\documentclass[%
footinbib,
superscriptaddress,
twocolumn,
amsmath,amssymb,
aps,
prl
]{revtex4-2}

\usepackage{ bbold }
\usepackage{graphicx}
\usepackage{dcolumn}
\usepackage{bm}
\usepackage[justification=RaggedRight]{caption}
\usepackage{subcaption}
\usepackage[utf8]{inputenc}
\usepackage[T1]{fontenc}
\usepackage{braket}
\usepackage{soul} 

\usepackage{dsfont}

\usepackage{xcolor}		
\usepackage[english]{babel} 

\newbox{\bigpicturebox}

\usepackage{csquotes}
\usepackage{hyperref}
\usepackage{bookmark}

\hypersetup{%
  breaklinks=true,
  colorlinks=true,
  linkcolor=black,
  filecolor=black,
  urlcolor=black,
  citecolor=black,
}

\begin{document}

\title{Non-reciprocal spin-glass transition and aging}

\author{Giulia Garcia Lorenzana}
\affiliation{Laboratoire de Physique de l'\'Ecole normale sup\'erieure, ENS, Universit\'e PSL, CNRS, Sorbonne Universit\'e, Universit\'e de Paris F-75005 Paris, France}
\affiliation{Laboratoire Matière et Systèmes Complexes (MSC), Université Paris Cité, CNRS, 75013 Paris, France}

\author{Ada Altieri}
\affiliation{Laboratoire Matière et Systèmes Complexes (MSC), Université Paris Cité, CNRS, 75013 Paris, France}
\author{Giulio Biroli}
\affiliation{Laboratoire de Physique de l'\'Ecole normale sup\'erieure, ENS, Universit\'e PSL, CNRS, Sorbonne Universit\'e, Universit\'e de Paris F-75005 Paris, France}
\author{Michel Fruchart}
\affiliation{Gulliver, ESPCI Paris, Université PSL, CNRS, 75005 Paris, France}
\author{Vincenzo Vitelli}
\affiliation{James Franck Institute, University of Chicago, Chicago, Illinois, 60637, U.S.A.}
\affiliation{ Department of Physics, University of Chicago, Chicago, Illinois, 60637, U.S.A.}
\affiliation{Kadanoff Center for Theoretical Physics, University of Chicago, Chicago, IL 60637, U.S.A.}

\begin{abstract}

Disordered systems generically exhibit aging and a glass transition. Previous studies have long suggested that non-reciprocity tends to destroy glassiness. Here, we show that this is not always the case using a bipartite spherical Sherrington-Kirpatrick model that describes the antagonistic coupling between two identical complex agents modelled as macroscopic spin glasses. Our dynamical mean field theory calculations
reveal an exceptional-point mediated transition from a static disorder phase to an oscillating amorphous phase as well as non-reciprocal aging with slow dynamics and oscillations.

\end{abstract}
\maketitle



%
Glassy systems do not reach equilibrium even on very long time scales~\cite{Houches2002,Keim2019,Arceri2021}.
The older the glassy system is, the slower it evolves: the typical relaxation time scale of a sample increases as the time elapsed since its preparation increases \cite{bouchaud1997a,biroli2005, cugliandolo2003}.
This very slow dynamics, known as \textit{aging},
has been observed in physical systems ranging from disordered magnets and spin-glasses to dense liquids and active matter~\cite{bouchaud1997a,biroli2005, cugliandolo2003,berthier2011theoretical,berthier2013, keta_disordered_2022, janssen_active_2019, berthier_glassy_2019, ghimenti_transverse_2024, ghimenti_irreversible_2024,Ghimenti2023,Mandal2020}.
Glass-like dynamics have also been observed in ecological systems~\cite{altieri2021,altieri2022,Altieri2022b,loreau2013,allesina2012, galla2006,ros2023a,arnoulx2024many} as well as networks of biological or artificial neurons~\cite{Hopfield1982,Hertz1986,dayan2001, sompolinsky1988, parisi1986, martorell2023,derrida1987,aguirre-lopez2022,Clark2024,Marti2018,Dahmen2019,Biroli2023,Biroli2024,BaityJesi2018a}.
Notably, these systems can exhibit non-reciprocal interactions between constituents (think of predator-prey relationships in ecology) and therefore need not satisfy micro-reversibility.

Understanding how the dynamics of non-reciprocal systems becomes glassy is a major challenge that goes beyond the body of literature developed in the last decades on glassy systems~\cite{mezard1986,Charbonneau2023}. 
Pioneering studies by Crisanti and Sompolinsky (CS)~\cite{crisanti1987} considered a spherical Sherrington-Kirkpatrick (SK) model \cite{Sherrington1975,Parisi1979,cugliandolo1995, kosterlitz1976, dedominicis2006} to which they added random {\it all-to-all} non-reciprocal interactions between spins. 
They showed that non-reciprocity suppresses the finite temperature spin-glass transition and, hence, also aging dynamics. Instead, chaotic dynamics is observed.
These results were extended to more general glassy models in Refs.~\cite{cugliandolo1997glassy,berthier2000, horner_drift_1996,fyodorov2023non} and are believed to originate from the marginal stability of the model (i.e. the presence of flat directions in phase space \cite{muller2015marginal}).
The emerging picture is that any amount of non-reciprocal interactions then tends to destroy glassiness \footnote{This is not the case for discrete models, where often a finite amount of non-reciprocity is needed to destroy aging behavior \cite{Iori1997,garnier-brun2024}}.

In this Letter, we demonstrate that the aforementioned conclusions depend on the topology of the network of non-reciprocal interactions, and their distribution. In  particular, our analysis accounts for the scenario in which \textit{agents} that are coupled non-reciprocally are themselves macroscopic entities with complex internal dynamics. Examples include predators and preys, adversarial neural networks or robots, etc.
Specifically, we consider a bipartite many-body spin-glass system with random symmetric interactions in each part, and fixed antisymmetric interactions between the two parts.
We find evidence of (i) a finite temperature non-reciprocal spin-glass phase transition between a static disordered phase and a time-dependent amorphous phase and (ii) non reciprocal aging characterized by both slow dynamics and oscillations. 
The mechanism underpinning the destabilization of the usual spin glass in favor of the non-reciprocal one is a spectral singularity called exceptional point.

Non-reciprocal interactions considered here have been studied in a variety of contexts where they generate a rich phenomenology ranging from oscillatory states and traveling waves \cite{avni2025a, guislain2024a, guislain2023a, guislain2023, martin2023,fruchart2021,Saha2020,You2020,loos2023} to chaotic states \cite{crisanti1987, dinelli2023, martorell2023, sompolinsky1988, zakine_socioeconomic_2024, parisi1986} and nonequilibrium phase transitions~\cite{fruchart2021,Hanai2024,Daviet2024,Zelle2024,avni2025a, ottino-loffler2018, pruser2024a, pruser2024,Daido1992,Daido1987, garces2024, garnier-brun2024}. 
In neural networks, asymmetric interactions also play a role in learning processes~\cite{Amit1989,Lecun1985,parisi1986,fasoli2018,sompolinsky1988,Sompolinsky1986,Rabinovich2006,Khona2022}.
In the context of spin glasses, it has been shown in Ref.~\cite{guislain2024b} that the oscillatory dynamics often encountered in this class of systems persists in mean-field Mattis-like models in which disorder can be gauged away by a change of variables \cite{Mattis1976,Toulouse1987}.
However, these systems are not marginal, leaving open the question of how non-reciprocity affects marginally stable glassy systems.
Several extensions of our model are discussed in the accompanying publication \cite{garcia_lorenzana_nonreciprocally_2025}.

 
\begin{figure}
\centering
\includegraphics[width=0.27\textwidth]{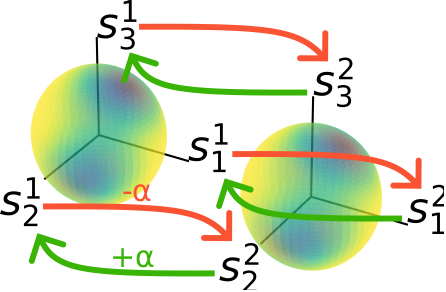}
\caption{Sketch of the non-reciprocal spin-glass model: two ($N$-dimensional, in the plot $N=3$) spherical spin systems, corresponding degrees of freedom are coupled non-reciprocally. The sphere's colors sketch the identical random potential in both systems.}
\label{fig:sketchSK}
\end{figure}

\textit{Model.---} To construct a minimal model exhibiting marginal glassy dynamics in the absence of non-reciprocity, we follow the route of Crisanti and Sompolinsky but with a crucial difference. We consider {\it two} spherical (SK) spin-glass systems each composed of $N$ spins representing two distinct species denoted by $1$ and $2$ with random all-to-all symmetric interactions within the same species plus a non-reciprocal (antisymmetric) deterministic coupling between the two species. This setup is described by the Langevin dynamics:
\begin{align}
\label{eq:motion}
\begin{cases}                     
    \dot{s}^1_i=\sum_j J_{ij} s^1_j- \ell^1 s^1_i+ \alpha s^2_i + \eta^1_i + h_1\\   
    \dot{s}^2_i=\sum_j J_{ij} s^2_j- \ell^2 s^2_i- \alpha s^1_i + \eta^2_i + h_2
\end{cases}
\end{align}  
in which $s^a_i$ is the state of the spin $i=1,\dots,N$ in the system $a=1,2$.
The quenched interaction matrix $J_{ij}$ is symmetrical and assumed to be the same in both systems (see \cite{garcia_lorenzana_nonreciprocally_2025} for different interaction matrices).
The elements of $J_{ij}$ are drawn from a Gaussian distribution with zero mean and variance $\sigma^2/N$ ($\sigma=1$ in the following). 
Corresponding spins in each system $s^1_i$ and $s^2_i$ are coupled antisymmetrically, with a coupling strength $\alpha$: system 1 wants to align with system 2,  whereas system 2 wants to anti-align with system 1. These non-reciprocal interactions correspond to antagonistic attraction and repulsion between 1 and 2.  See Figure \ref{fig:sketchSK} for a graphical sketch.
Each spin is also subject to a thermal Gaussian white noise $\eta_i^a$ with covariance $\langle \eta_i^a(t) \eta_j^b(t')\rangle=2T\delta_{ij}\delta_{ab}\delta(t-t')$, where $T$ denotes the temperature. 
$h_1$ and $h_2$ are external fields that will be taken to zero at the end of the computation.
Finally, $\ell^1$ and $\ell^2$ are Lagrange multipliers used to enforce the spherical constraints $\frac{1}{N}\sum_{i=1}^N(s_i^a)^2=1$. 
This model can be interpreted as describing the dynamics of two identical agents 
that are antagonistically coupled and whose internal complexity is modeled by spin-glass dynamics.
Crisanti and Sompolinsky considered instead a single spherical SK model, with non symmetric interaction matrix $J_{ij}$.

\textit{Non-reciprocal spin-glass transition ---}
Without non-reciprocity, the spherical spin-glass model displays a finite temperature phase transition between a high-temperature disordered phase and a low-temperature ordered phase. The latter is characterized by two ground states related by spin-inversion symmetry (in replica language, it reflects a simple replica symmetric spin-glass phase) \cite{cugliandolo1995, dedominicis2006}.  The transition can also be detected dynamically: starting from random initial conditions, at high temperature the system converges to a steady state, whose relaxation time diverges when approaching $T_c$.  In the following, we analyze the effect of non-reciprocity on the steady-state dynamics by progressively decreasing the temperature. 

In the thermodynamic limit $N\to\infty$, the dynamics of the system can be analyzed by Dynamical Mean-Field Theory (DMFT) \cite{crisanti1987, sompolinsky1982, ABC2020, cugliandolo2023}.  DMFT allows to obtain effective equations for a given couple $s_i^1$, $s_i^2$, in which interactions with other spins are replaced by additional noise and memory terms. Their statistics have to be computed in a self-consistent way (see \cite{garcia_lorenzana_nonreciprocally_2025} for the derivation).
Since the resulting equation does not depend on $i$, we drop the index $i$ in the following and rewrite the equation in vectorial form as:
	\begin{align}
    \label{eq:dmft}
		\dot{\mathbf{s}} = - \Lambda \mathbf{s}+ \alpha\epsilon \mathbf{s}+\mathbf{\xi}+\eta +\int_0^t dt'R(t,t')\mathbf{s}(t') +\mathbf{h} \ ,
\end{align}
where 
$\mathbf{s} = [s^1, s^2]^T$
is a two-dimensional vector that contains the two spins, $\Lambda=\text{diag}(\ell^1, \ell^2)$ is the diagonal matrix of the Lagrange multipliers, 
$\epsilon$ is the fully anti-symmetric Levi-Civita symbol, and $\eta = [\eta^1, \eta^2]^T$.  The self-consistent noise vector $\mathbf{\xi}$ is Gaussian with zero mean and variance $\langle \xi_a(t)\xi_b(t')\rangle = C_{ab}(t, t') $. 
Finally,
\begin{align}
    C_{ab}(t, t')= \langle s_a(t)s_b(t')\rangle , &&    R_{ab}(t,t')= \frac{\partial\langle s_a(t)\rangle}{\partial h_b(t')} \ .
\end{align}
are the average correlation and response matrices, which have to be determined self-consistently.
At high temperature, the system reaches a time-translation invariant (TTI) state (see the End Matter). We can then compactly write the self-consistent equations for the correlation $C$ and the response $R$ in Fourier space:
\begin{align}
	R^{-1}(\omega)&=(-i\omega+\ell) \mathbb{1} -R(\omega) -\alpha \epsilon\label{eq:Rselfconsistent}\\
    C(\omega) &= 2 T ((R(\omega)^\dag R(\omega))^{-1}-\mathbb{1})^{-1}  \ ,\label{eq:Cselfconsistent}
\end{align}
where $\ell$ is the value of the two Lagrange multipliers, which are constant and equal in the TTI regime.
Note that because the system is symmetric under the transformation $s_1\to -s_2$, $s_2\to s_1$, both $R$ and $C$ have only two independent elements: the diagonal elements are equal ($R_{11}=R_{22}=R_d$), whereas the off-diagonal ones are equal and opposite ($R_{12}=-R_{21}=R_a$).
Also, because of the spherical constraint, the autocorrelation function must equal 1 for $t=0$. 

	\begin{figure*}[htbp]
		\centering 
     \begin{subfigure}[b]{0.3\textwidth}
         \centering
		\includegraphics[width=\textwidth]{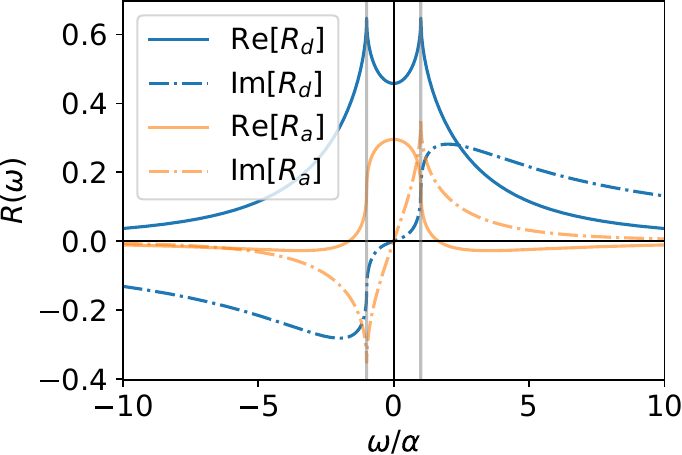}
         \caption{Response function}
         \label{fig:Romega}
     \end{subfigure}
     \hfill
     \begin{subfigure}[b]{0.3\textwidth}
         \centering
      \includegraphics[width=\textwidth]{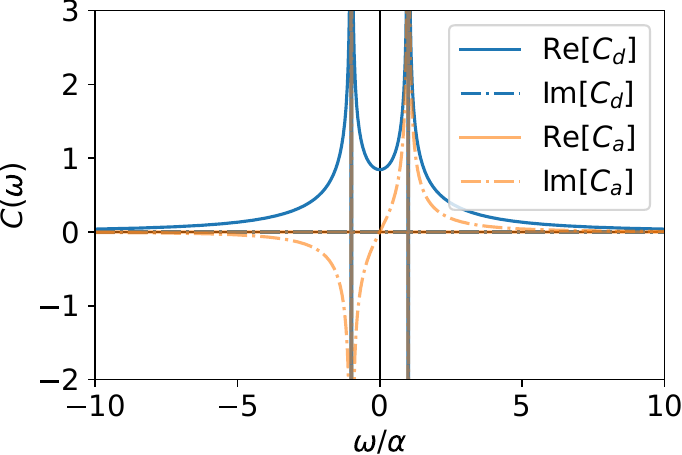}
         \caption{Correlation function}
         \label{fig:Comega}
     \end{subfigure}
     \hfill
         \begin{subfigure}[b]{0.32\textwidth}
         \centering
      \includegraphics[width=\textwidth]{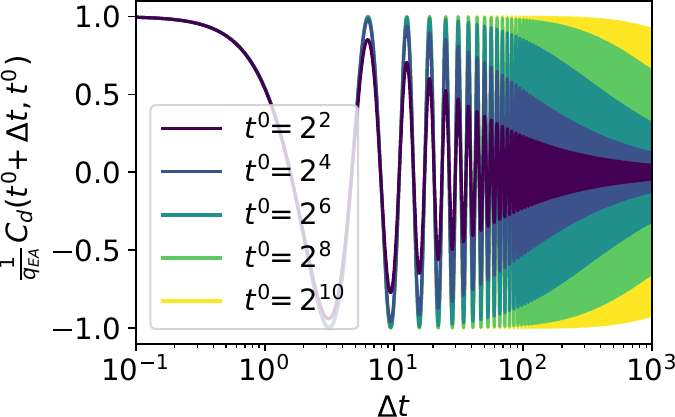}
         \caption{Aging correlation function}
         \label{fig:Caging}
     \end{subfigure}
\caption{Real and imaginary part of the diagonal and antisymmetric components of the response (a) and correlation (b) functions at the critical point. In gray we highlight $\omega/\alpha=\pm 1$. $T=T_c=1$. (c) Diagonal component of the correlation function, normalized by the Edward-Anderson order parameter, for different values of the waiting time. 
}
  \label{fig:varieRC}
	\end{figure*}

 By solving these equations, we find that both response and correlation functions are peaked around $\omega=\pm \alpha$ (Fig. \ref{fig:Romega}-\ref{fig:Comega} and \cite{garcia_lorenzana_nonreciprocally_2025}).
 This is the frequency at which the system would exhibit regular oscillations in the absence of disordered interactions. 
Strikingly, in their presence, the disordered system does not exhibit macroscopic oscillations but responds more strongly if excited at this frequency.
 These peaks lead to bona-fide singularities when $T\to T_c=1$ (Figure \ref{fig:Romega}-\ref{fig:Comega}): the one of $R(\omega)$ approaches a finite limit with a square root behavior, whereas the one of $C(\omega)$ diverges as $(\omega \mp\alpha)^{-1/2}$.  In the time domain,  one finds a relaxation time to the non-equilibrium steady state that diverges as $1/\sqrt{T-T_c}$ and critical relaxation at $T_c$, corresponding to a behavior $C_d(t)\sim \cos(\alpha t)/t^{1/2}$.  

This transition shares crucial similarities to the transition to an ordinary spin glass phase that is found in the reciprocal (uncoupled) case except for the superposition of oscillations that shift the singularity of correlation and response functions from $\omega=0$ to $\omega=\pm\alpha$.
The self-consistency equations can be expressed in terms of the eigenvalues of $R$ and $C$, and in this form they can then be shown (EM) to reduce exactly to the ones obtained in the uncoupled case, except for a shift of $+\alpha$ or $-\alpha$ in the $\omega$ dependence of the two eigenvalues.  This mapping explains why the critical behavior and even the critical point $T_c=1$ are the same as in the uncoupled case.  The analysis of the steady state dynamics therefore reveals a first important result: adding non-reciprocal interactions to the spherical spin-glass model does lead to a dynamical phase transition,  which generalizes the spin-glass transition found in the symmetric case.  This is in contrast with what is found for all-to-all non-reciprocal interactions \`a la CS, which instead wipe out the finite temperature transition \cite{crisanti1987}. We will now show that the different form of non-reciprocity also leads to a very different physical behavior for the non-equilibrium dynamics.
 
 
\textit{Non-reciprocal aging  ---}
Since the relaxation time diverges at $T_c$, the dynamics after quenches below $T_c$ are not expected to relax to a steady state.  As we show below, indeed they do not --- instead aging ensues (on timescales that do not diverge with $N$). 
To study the aging regime we follow Ref.~\cite{cugliandolo1995} and analyze the dynamics in the basis that diagonalizes the interaction matrix $J$. This leads to
\begin{align}
\label{eq:smu}
	\dot{\mathbf{s}}_\mu = \left( \mu\mathbb{1} - \Lambda + \alpha \epsilon \right)\mathbf{s}_\mu +\mathbf{\eta}_\mu
\end{align}
in which the eigenvalues $\mu$ of $J$ are used to label the two-dimensional vectors $\mathbf{s}^\mu$ composed of the projections of $s^1$ and $s^2$ on the corresponding eigenvector.
The different modes are now coupled only through the Lagrange multipliers, which will now be time-dependent. 
At $t=0$ each mode is initialized as a Gaussian random variable with mean zero and variance one, corresponding to a sudden quench from infinite temperature to the temperature $T$. Using a procedure similar to Ref.~\cite{cugliandolo1995}
we find that after the quench $\mathbf{s}_\mu$ rotates at constant angular velocity $\alpha$, while its radius undergoes a slow aging evolution akin to the one found in the absence of non-reciprocity (EM). 
This decoupling of the oscillatory and aging dynamics is a non-trivial result of our analysis. 
The self-correlation functions, $C_{11}$ and $C_{22}$, plotted in Figure \ref{fig:Caging} for different values of the first time $t'$,  read in the asymptotic regime $t, t'\gg 1$, $\Delta t= t-t'\gg 1$:
\begin{align}
    C_{d}(t, t')= q_{EA}\left(\frac{2 \sqrt{1+\Delta t/t'}}{2+\Delta t/t'}\right)^{\frac{3}{2}} 
    \cos(\alpha \Delta t)
\end{align} 
where $q_{EA}=1-\frac{T}{T_c}$ is the non-reciprocal counterpart of the Edwards-Anderson order parameter or self-overlap \cite{cugliandolo1995, dedominicis2006, edwards1975, parisi2004} (see EM for the expression for $C_{a}(t,t')$).
This form is in very good agreement with the simulation results (EM).
In summary, in the non-reciprocal case, the correlation function after a quench displays both an oscillating behavior due to non-reciprocal interactions and a slow aging evolution -- with a dependence on $\Delta t/t'$, which remarkably turns out to be the same as in the reciprocal case \cite{cugliandolo1995, dedominicis2006}.
The dynamics of the system is therefore controlled by two independent timescales: one, proportional to $1/\alpha$, sets the period of the oscillations, whereas the other, proportional to the \textit{age} of the system $t'$, determines the coherence time of the oscillations. We call this phenomenon \textit{non-reciprocal aging}.
Notably, aging survives for all values of the non-reciprocity $\alpha$, contrarily to previously studied cases in which aging is interrupted above some critical value of the driving \cite{Berthier2001, Iori1997, garnier-brun2024}.
It is instead necessary for the two systems to have the same interaction matrix, otherwise the aging dynamics acquires a finite lifetime \cite{garcia_lorenzana_nonreciprocally_2025}.

\begin{figure}
\centering
\includegraphics[width=0.3\textwidth]{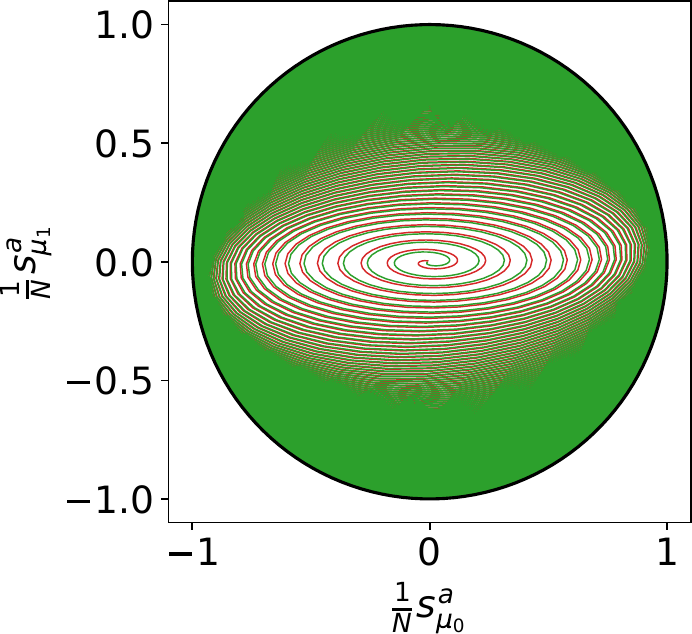}
\caption{Projection on the two leading eigenvectors of the interaction matrix $J$ of the trajectories (green and red) of the two systems. Random initial conditions, $T=0$, $\alpha=0.2$, $N=20000$, $t_{max}=2000$.}
\label{fig:orbit}
\end{figure}

\begin{figure}
\includegraphics[width=0.45\textwidth]{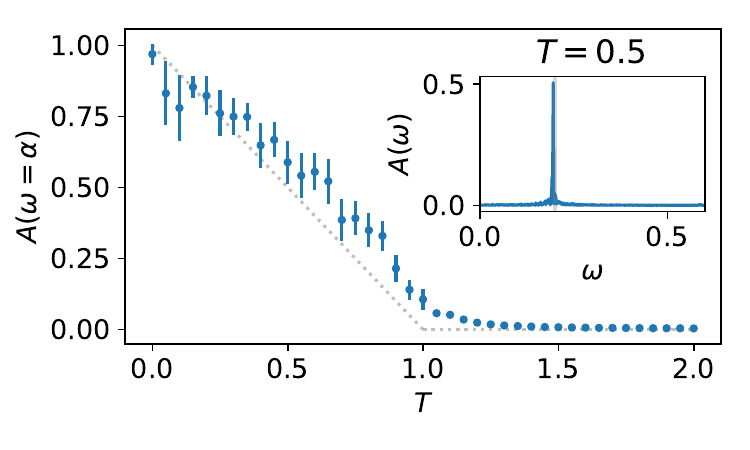}
\caption{Numerical results for the amplitude of the Fourier component at $\omega=\alpha$ of the projection of one of the two clones on the leading eigenvector, $\frac{1}{N}\sum s_i^a(v_{\mu_0})_i$, as a function of temperature. The amplitude goes to zero continuously for $T\to T_c=1$; in grey we show $q_{EA}=1-\frac{T}{T_c}$. Inset: amplitude of the Fourier transform as a function of $\omega$ for $T=0.5$. $\alpha=0.2$}
\label{fig:FouriervsT}
\end{figure}

\textit{Asymptotic behavior ---}
At finite $N$, the relaxation timescale is finite. 
Therefore, if we consider $t\rightarrow \infty$ at large but fixed $N$, the system stops aging and we can access a different asymptotic regime.
In the $\alpha=0$ case, the system equilibrates to one of the two pure states associated with spin configurations oriented in the direction of the leading eigenvector of $J$, denoted as $v_{\mu_0}$, and then on time-scales exponentially large in $N^{1/3}$ \cite{rodgers1989, barbier2021a} the system switches from one state to the other by activated barrier hopping. By numerically integrating the equations of motion (\ref{eq:motion}), we find that the situation changes drastically once $\alpha$ is switched on. Activated barrier hopping is wiped out by the non-reciprocity and each of the systems oscillates on timescales of order one between the two states.   
More precisely, at zero temperature each system performs a periodic orbit in the circle spanned by the two lowest eigenvalues of $J$ (Figure \ref{fig:orbit}) with angular frequency $\alpha$ and relative phase $\pi/2$. 
At finite temperature, the Fourier transform of the projection of the spin configuration of any of the two systems on the leading eigenvector has a delta peak in $\omega=\alpha$, whose amplitude $A(\omega=\alpha)$ goes to zero continuously at the transition (Figure \ref{fig:FouriervsT}).
This oscillating phase, which we call non-reciprocal spin-glass, is the counterpart of the low-temperature static phase found at equilibrium. 
Because the oscillations are macroscopic, they are unaffected by thermal fluctuations, and the direction of rotation is randomly selected but never inverted in the numerically accessible times \footnote{As in the equilibrium case, we expect this inversion to occur on time-scales exponentially large in a power of $N$}.
Note that these persistent oscillations are a non-trivial effect of random interactions, that allow the system to condensate on a low-dimensional manifold: without them oscillations would be exponentially damped at any finite temperature \cite{garcia_lorenzana_nonreciprocally_2025}.
The observable $A(\omega=\alpha)$ generalizes the Edwards-Anderson parameter $q_{EA}$ that in the uncoupled case describes the overlap of the system with the ground state.
The numerical dependence on temperature of $A(\omega=\alpha)$ indicates a phase transition at $T_c=1$, and is quantitatively similar to the one of $q_{EA}$, $A(\omega=\alpha)=1-\frac{T}{T_c}$.

\textit{Exceptional-point mediated transition ---}
To investigate analytically the existence of the oscillating phase described above, we now analyze the stability of the equilibrium points at zero temperature. 
In the presence of non-reciprocal forces, $\pm \sqrt{N}v_{\mu_0}$ are still equilibrium points for both systems. We study their stability by linearizing the dynamics \eqref{eq:smu} around them (we stick to the case $s^1=s^2= \sqrt{N}v_{\mu_0}$ for simplicity). 
The stability matrix is block diagonal in the basis that diagonalizes $J$. 
For each $\mu$ the corresponding block in the stability matrix reads (EM):
 \begin{align}
    M_{\mu}= \begin{pmatrix}
 \mu- \mu_0-\alpha & \alpha \\
-\alpha & \mu-\mu_0 + \alpha 
\end{pmatrix}
\end{align}
The matrix $M_\mu$ is not diagonalizable, because it has only one eigenvector $(1,1)^\top$ (instead of two).
This situation is known as an exceptional point, a spectral singularity that is associated to the coalescence of two or more eigenstates, and that also arises in dissipative quantum systems\cite{bergholtz_exceptional_2021,Kato1984}. 
The unique eigenvalue $\lambda=\mu-\mu_0$ is always positive because $\mu_0$ is the lowest eigenvalue of $J$. 
Nevertheless, for non-normal matrices (such as those close to or at an exceptional point), looking at the eigenvalues of the stability matrix is not sufficient to determine the behavior of the system around the equilibrium point \cite{trefethen2005}. 
We find by solving the linearized dynamics (EM) that a perturbation on the mode $\mu$ is initially amplified as long as $\mu-\mu_0<\alpha$, although $M_\mu$ has only non-negative eigenvalues.
In the thermodynamic limit, the gap between the first and subsequent eigenvalues vanishes, therefore for any finite value of $\alpha$ there will be an extensive number of unstable modes destabilizing the ground state.
The destabilization is dominated by the most unstable mode, corresponding to the second lowest eigenvalue of $J$, and leads to the rotation in the circle spanned by the two lowest eigenvalues described before. 
Perturbations around this rotating state are exponentially damped, except for the zero mode corresponding to shifting the phase of the rotation \cite{garcia_lorenzana_nonreciprocally_2025}. 
This exceptional-point mediated transition is reminiscent of the non-reciprocal phase transitions studied in Ref.~\cite{fruchart2021}.
Our results provide an extension of that mechanism to simple disordered systems. 
As in \cite{fruchart2021}, at finite temperature the oscillating phase is also found if the interactions between the two systems are not exactly antisymmetric \cite{garcia_lorenzana_nonreciprocally_2025}.

In the equilibrium case, there is a symmetry breaking corresponding to the choice of one of the two pure states. 
In the non-reciprocal case, there is spontaneous chiral symmetry breaking: the direction of rotation is randomly selected, and with a rotation plane that is disorder-dependent.
As a result, similarly to what was found in references \cite{guislain2024,guislain2024b} for a related model, the oscillations are not visible in the magnetization of the system, that is zero in this phase, or other disorder-independent one-time observables. 
Instead, they are visible in the auto-correlation function \footnote{The plane of rotation can be obtained via principal component analysis (PCA) of the trajectory of the system.}.

\textit{Random non-reciprocity: continuous vs. discrete distributions. ---}
We emphasize that our non-reciprocal coupling $\alpha$ introduces a unique timescale for oscillations, in contrast with the random non-reciprocity in CS model, which leads to a continuous timescale distribution.
For continuous distributions of random anti-symmetrical coupling $\alpha_i$ \cite{garcia_lorenzana_nonreciprocally_2025}, we find that the transition is suppressed at any finite temperature, destroying the aging behavior and leading to chaotic dynamics as in the CS model \footnote{This is reminiscent of what was found for slowly pulsating poly-disperse particles \cite{Tjhung2017}, that exhibit an arrested phase only if all particles pulsate at the same frequency.}.
We have also considered anti-symmetric couplings which can only assume two different values with different probabilities. In this case, a genuine transition still exists. This raises the intriguing possibility of two different scenarios for non-reciprocal interactions: (i) one characteristic of continuous distributions first evidenced by Crisanti and Sompolinsky, (ii) one characteristic of discrete distributions, whose simplest incarnation is the system analyzed in this paper. Physically, continuous vs discrete distributions indeed correspond to quite different situations. The latter can be seen as macroscopic sub-systems coupled with different non-reciprocal interactions.   

To sum up, by studying minimal spin-glass models, we have shown 
that non-reciprocal interactions between two distinct species can lead to an exceptional-point mediated spin-glass phase and a novel mechanism of non-reciprocal aging with potential implications for many-body systems in which complex agents with antagonistic goals are themselves modeled as macroscopic disordered systems.
Possible extensions include generalizations to non-linear spin-glass models. 

\begin{acknowledgments}
We thank Y. Fyodorov for interesting discussions. This work was supported by the Simons Foundation Grant No. 454935 (G.B.). A.A. acknowledges the support received from the Agence Nationale de la Recherche
(ANR) of the French government, under the grant ANR-23-CE30-0012-01 (SIDECAR project). M.F. acknowledges partial support from the National Science Foundation under grant DMR-2118415, a Kadanoff–Rice fellowship funded by the National Science Foundation under award no. DMR-2011854 and the Simons Foundation.
V.V. acknowledges partial support from the Army Research Office under grant  W911NF-22-2-0109 and W911NF-23-1-0212 and the Chan Zuckerberg Initiative.
M.F. and V.V acknowledge partial support from the France Chicago center through a FACCTS grant. 
This research was partly supported from the National Science Foundation through the Center for Living Systems (grant no. 2317138) and the National Institute for Theory and Mathematics in Biology (NITMB).
\end{acknowledgments}

\bibliography{MyLibrary}

\appendix

 \section{Equality of Lagrange multipliers}
 The Lagrange multipliers correspond to the force that the constraints need to apply on the systems in order to keep them on the sphere. 
Because of the symmetry of the model under the transformation $s_1\to -s_2$, $s_2\to s_1$ (and under inversion) the two systems are completely equivalent. 
Whenever this equivalence is not broken, the force on the constraint is the same in the two systems, and therefore the two Lagrange multipliers are equal, $\ell^1=\ell^2= \ell$.
We expect this to be true as long as an extensive number of modes contribute to the dynamics, allowing us to interpret the sum of the force over modes as an average. 


\section{Time-translational-invariant phase}

\begin{figure}[htbp]
    \centering
    \includegraphics[width=0.3\textwidth]{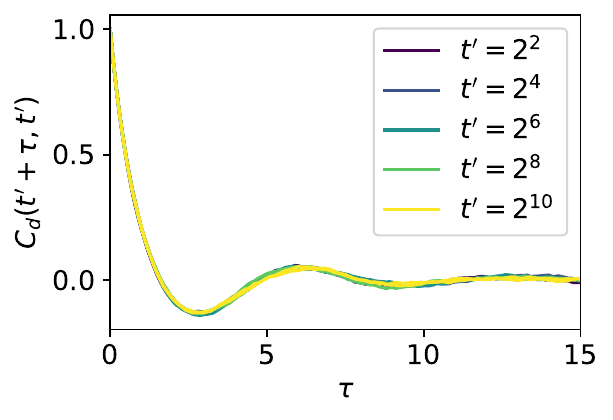}
    \includegraphics[width=0.3\textwidth]{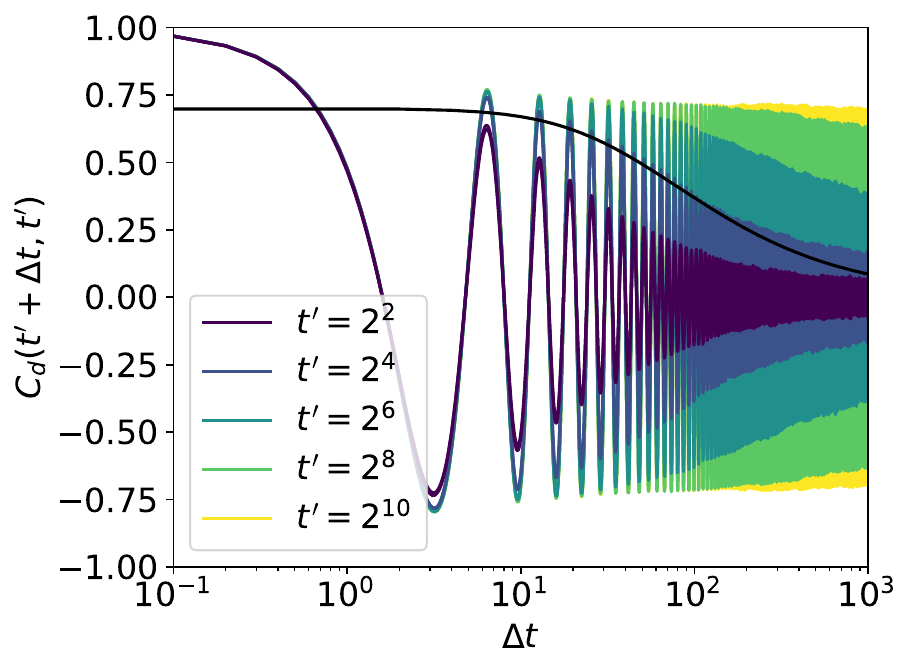}
    \caption{Correlation functions for different values of the initial time $t'$, from numerical simulations of the equations (\ref{eq:motion}), above (top, $T=1.4$) and below (bottom, $T=0.3$) the critical point. $\alpha=1$, $N=20000$, averaged over 5 runs. In black the analytical prediction for the envelope of the oscillations (i.e. $C_0(t, t')$) for $t'=2^4$.}
    \label{fig:Cnum}
\end{figure}
In Fig. \ref{fig:Cnum} we show the auto-correlation function of the system $C(t'+\tau, t')$ for different values of the initial time $t_0$, obtained from numerical simulations of equations (\ref{eq:motion}). For $T>T_c$ (top), it only depends on the time difference $\tau$, confirming that the system is in a time translational invariant state.
When this is the case, the DMFT equations greatly simplify in Fourier transform:
\begin{align}
    ((-i\omega+\ell) \mathbb{1} -R -\alpha\epsilon)\mathbf{s}= \mathbf{\xi}+\mathbf{\eta} + \mathbf{h}
\end{align}
Differentiating both sides with respect to $\mathbf{h}$ we obtain Eq. \eqref{eq:Rselfconsistent}.
If we knew the Lagrange multiplier $\ell$, this equation would determine $R$. 
Because of the aforementioned symmetry under the transformation $s_1\to -s_2$, $s_2\to s_1$, $R$ has the form 
    $R=  \begin{pmatrix} R_d  & R_a\\  -R_a & R_d \end{pmatrix}$. 
All matrices of this form are diagonalized in the basis $v^\pm =\frac{1}{\sqrt{2}} \begin{pmatrix} \mp i \\   1 \end{pmatrix}$. 
We can write the self-consistent equation on $R$ in its eigenbasis:
\begin{align}
    \ell- i (\omega \pm \alpha) - R_\pm = R_\pm^{-1}
\end{align}
As noted in the main text, this is the same result as in the uncoupled case, except for a shift of $\pm \alpha$ in $\omega$.


Using that $\mathbf{s}(\omega)=R(\omega)(\mathbf{\xi}(\omega)+\mathbf{\eta}(\omega))$, we can obtain a self-consistent equation on $C$:
\begin{align}
     C(\omega) =  R(\omega)( C(\omega)+2T) R^\dag(\omega)    
\end{align}

As $R$, also $C$ is of the form
$ C =  \begin{pmatrix} C_d & C_a\\  -C_a & C_d \end{pmatrix}$. 
This means that it is diagonalized in the same basis $v^\pm$ as $R$, which ensures that $C$ and $R$ commute. If $(R(\omega)^\dag R(\omega))^{-1}-\mathbb{1}$ is invertible, we then obtain Eq. \eqref{eq:Cselfconsistent}. 
In the diagonalizing basis, we have:
\begin{align}
    C_\pm(\omega)  = \frac{2T}{|R_\pm(\omega)|^{-2}- 1}
\end{align}
This is again the same equation as in the uncoupled case. Because the $\omega$ dependence is only through $R_\pm$, also $C$ behaves in the same way as in the uncoupled case except for the $\pm \alpha$ shift in $\omega$. 

The Lagrange multiplier $\ell$ is determined by imposing the spherical constraint, $C_d(t=0)=1$. Since 
\begin{align}
\begin{split}
\label{eq:sphericalconstraintC}
    \int d\omega C_d(\omega)=\int d\omega C_\pm(\omega) =\int d\omega C_{\alpha=0}(\omega) 
\end{split}
\end{align}
 $\ell$ will be at all temperatures the same as in the uncoupled case. 
This leads to the same critical point $T_c=1$, where $\ell\to 2$ and touches the edge of the spectrum of $J$. 

For $\ell=2$ one eigenvalue of $R$ behaves around $\omega=\alpha$ as $1- \sqrt{|\omega-\alpha|}$. The corresponding eigenvalue of $C$ therefore behaves as $2T|\omega-\alpha|^{-1/2}$, leading to an integrable singularity.
If this was not the case (as for example in reference \cite{crisanti1987}), the integral in equation (\ref{eq:sphericalconstraintC}) would be diverging at criticality, implying that criticality can only be reached at 0 temperature.

\section{Aging}

In the basis that diagonalizes the interaction matrix $J$ (Eq. \eqref{eq:smu}), different modes are only coupled through the Lagrange multipliers. 
We can then formally write the solution in terms of the noise $\eta_\mu(t)$ and the (unknown) time evolution of the Lagrange multipliers:
\begin{align}
\begin{split}
    \mathbf{s}_\mu(t) =e^{\left( \mu \mathbb{1}+ \alpha\epsilon \right)t- \int_0^t\Lambda(t')dt' }\mathbf{s}_\mu (0)+\\+\int_0^tdt'e^{\left( \mu \mathbb{1}+ \alpha\epsilon \right)(t-t')- \int_{t'}^t\Lambda(t'')dt'' }\eta_\mu(t')    
\end{split}
\end{align}

On time scales that do not diverge with $N$, an extensive number of modes will contribute to the dynamics of the Lagrange multipliers. 
As before, and because of the randomness in the initial conditions, we then expect them to be equal at all times, $\ell^1(t)=\ell^2(t)=\ell(t)$. 
This greatly simplifies the analysis, because now $\Lambda$ and $\epsilon$ commute and we can separate their exponentials:
\begin{align}
\begin{split}
    \mathbf{s}_\mu(t) =e^{\mu t- \int_0^t\ell(t')dt' }R_{\alpha t} \mathbf{s}_\mu (0)+\\+\int_0^tdt'e^{\mu (t-t')- \int_{t'}^t\ell(t'')dt'' }R_{\alpha (t-t')}\eta_\mu(t') 
\end{split}
\end{align}
where $ R_{\alpha t} = e^{\alpha\epsilon t} = \begin{pmatrix} \cos \alpha t & \sin \alpha t  \\ - \sin \alpha t & \cos\alpha t \end{pmatrix}$. 
Imposing the spherical constraint and averaging over uniform initial conditions on the two spheres we get an equation on $\ell$:
\begin{align}
\label{eq:sphericalaging}
\begin{split}
    \int d\mu \rho(\mu)e^{2\mu t- 2\int_0^t\ell(t')dt' } +\\+2 T  \int d\mu \rho(\mu) \int_0^tdt' e^{2 \mu (t-t')- 2 \int_{t'}^t\ell(\tau)d\tau }=1    
\end{split}
\end{align}
$\rho(\mu)$ is the eigenvalue density of $J$, given by the Wigner semicircle: $\rho(\mu)= \frac{1}{2\pi} \sqrt{4-\mu^2}$ for $\mu\in [-2,2]$.
Eq. \eqref{eq:sphericalaging} is the same equation that would be obtained in the uncoupled case \cite{cugliandolo1995}, therefore we can use the known result for the spherical constraint. 
This allows us to compute the correlation function, which turns out to be the same as in the uncoupled case, multiplied by the rotation matrix \cite{cugliandolo1995, garcia_lorenzana_nonreciprocally_2025}:
\begin{align}
    C(t,t')=C_{\alpha=0}(t, t')R_{\alpha (t-t')}
\end{align}


\section{Stability of equilibrium points}
\label{app:defectiveM}

In order to better understand the behavior of our system at long times, we can look at the fixed points of the deterministic dynamics ($T=0$). We use again the basis that diagonalizes the interaction matrix $J$:
\begin{align}
\label{eq:smufixedpoint}
\begin{cases}
	\dot{s}^1_\mu = \left( \mu - \ell^1\right)s^1_\mu + \alpha s^2_\mu=0\\
 	\dot{s}^2_\mu = \left( \mu - \ell^2\right)s^2_\mu - \alpha s^1_\mu=0    
\end{cases}
\end{align}
We have $4N$ stationary points; in each only one of the modes $\mu^*$ contributes: $s^1_{\mu}=\pm s^2_{\mu}=\pm\sqrt{N}\delta_{\mu, \mu^*}$.
Let us consider a fixed point $\mathbf{s}^*$ with positive projections on the mode $\mu^*$ for both clones (the other equilibria are equivalent thanks to symmetry). 
This determines the Lagrange multipliers: $\ell^1=\mu^*+\alpha$ and $\ell^2=\mu^*-\alpha$.

To study the stability of this fixed point we can linearize the dynamics, defining $\delta\mathbf{s}_\mu= \mathbf{s}_\mu-\mathbf{s}^*_\mu$:
\begin{align}
    \delta \dot{\mathbf{s}}_\mu &= M_\mu\delta \mathbf{s}_\mu= \begin{pmatrix}
 \mu- \mu^*-\alpha & \alpha \\
-\alpha & \mu-\mu^* + \alpha 
\end{pmatrix}\mathbf{s}_\mu
\end{align}
The stability matrix $M_\mu$ has only one eigenvalue $\lambda_\mu=\mu-\mu^*=\Delta \mu$, with only one associated independent eigenvector: it is a \textit{defective} matrix.
If $\mu^*$ is not the maximum eigenvalue, a finite number of the stability eigenvalues $\lambda_\mu$ will be positive, and therefore the system will depart exponentially from the fixed point. 
If $\mu^*$ is the maximum eigenvalue, all the stability eigenvalues will be negative, except for the one associated with mode $\mu^*$ which will be zero.
This is exactly what happens in the uncoupled case.
Nevertheless, because here $M_\mu$ is defective, even though all eigenvalues are non negative we are not guaranteed that a small perturbation around the fixed point will be exponentially damped.
Indeed, we can explicitly solve the linearized equation:
\begin{align*}
    \delta \mathbf{s}_\mu =e^{M_\mu t}\delta \mathbf{s}_\mu(0)= \begin{pmatrix}
e^{\Delta \mu t }(1-\alpha t) & \alpha e^{\Delta \mu t } t\\
-\alpha e^{\Delta \mu t } t & e^{\Delta \mu t }(1+\alpha t)
\end{pmatrix} \delta \mathbf{s}_\mu(0)
\end{align*}
Even though at long times the behavior is controlled by the decaying exponential, at short times we can see a growth of the perturbation. Expanding the exponential:
 \begin{align}
\delta \mathbf{s}_\mu \sim \begin{pmatrix}
1+(\Delta  \mu  -\alpha) t & \alpha  t\\
-\alpha  t & 1+(\Delta \mu  +\alpha )t
\end{pmatrix} \delta \mathbf{s}_\mu(0)    
\end{align}
Let us take for example the perturbation $\delta \mathbf{s}^\mu(0)= \varepsilon \begin{pmatrix} 0 \\1 \end{pmatrix}$. 
Because the two systems are aligned, while system 2 wants to antialign with system 1, we expect that a perturbation on system 2 could destabilize the system.
Indeed at short times we obtain $\delta \mathbf{s}^\mu(t)= \varepsilon \begin{pmatrix} \alpha  t \\1+(\Delta \mu  +\alpha )t \end{pmatrix}$. 

\section{Numerical simulations}

All numerical simulations are direct integrations of the corresponding Langevin equations by Euler–Maruyama method. The spherical constraints are imposed by normalizing each spherical system after each time step, $s_i^a=\tilde{s}_i^a/\left(\frac{1}{N}\sum_i (\tilde{s}_i^a)^2\right)$, where $\tilde{s}_i^a$ is the value obtained integrating the equations without taking into account the spherical constraint. This introduces an error of order $dt$, so that it does not change the precision of the method.
If needed, the Lagrange multipliers can be computed as the force that would be needed to keep the systems on the sphere:
$l_a=\frac{1}{dt}\left(\sqrt{\frac{1}{N}\sum_i (\tilde{s}_i^a)^2}-1\right)$.

\end{document}